%% file: main.tex
\UseRawInputEncoding

\RequirePackage{lineno} 
\documentclass[prd, reprint, preprintnumbers, twocolumn, eqsecnum,floatfix,letterpaper,superscriptaddress,nofootinbib]{revtex4-1}
\usepackage{graphicx,url,amssymb,amsmath,rotating,color,units,wasysym,epsfig,multirow,epstopdf,subcaption, booktabs, caption}
\usepackage{CJK}

\usepackage{booktabs}
\usepackage{dcolumn} 
\usepackage{colortbl}
\usepackage{xcolor}
\usepackage[colorlinks,urlcolor=blue,citecolor=blue,linkcolor=blue]{hyperref}
\usepackage{comment}
\usepackage{lineno}

\graphicspath{{./}{figures/}}

\begin{document}



\title{A machine learning-enabled search for binary black hole mergers in LIGO-Virgo-KAGRA's third observing run}

\author{Ethan Marx}
\affiliation{Department of Physics, MIT, Cambridge, MA 02139, USA}
\affiliation{LIGO Laboratory, 185 Albany St, MIT, Cambridge, MA 02139, USA}
\author{William Benoit}
\affiliation{School of Physics and Astronomy, University of Minnesota, Minneapolis, MN 55455, USA}
\collaboration{These authors contributed equally to this work}

\author{Trevor Blodgett}
\affiliation{Department of Physics and Astronomy, University of Toledo, Toledo, OH 43606, USA}
\author{Deep Chatterjee}
\affiliation{Department of Physics, MIT, Cambridge, MA 02139, USA}
\affiliation{LIGO Laboratory, 185 Albany St, MIT, Cambridge, MA 02139, USA}
\author{Emma de Bruin}
\affiliation{School of Physics and Astronomy, University of Minnesota, Minneapolis, MN 55455, USA}
\author{Steven Henderson}
\affiliation{School of Physics and Astronomy, University of Minnesota, Minneapolis, MN 55455, USA}
\author{Katrine Kompanets}
\affiliation{School of Physics and Astronomy, University of Minnesota, Minneapolis, MN 55455, USA}
\author{Siddharth Soni}
\affiliation{Department of Physics, MIT, Cambridge, MA 02139, USA}
\affiliation{LIGO Laboratory, 185 Albany St, MIT, Cambridge, MA 02139, USA}
\author{Michael Coughlin}
\affiliation{School of Physics and Astronomy, University of Minnesota, Minneapolis, MN 55455, USA}
\author{Philip Harris}
\affiliation{Department of Physics, MIT, Cambridge, MA 02139, USA}
\author{Erik Katsavounidis}
\affiliation{Department of Physics, MIT, Cambridge, MA 02139, USA}
\affiliation{LIGO Laboratory, 185 Albany St, MIT, Cambridge, MA 02139, USA}

\date{\today}

\begin{abstract}
We conduct a search for stellar-mass binary black hole mergers in gravitational-wave data collected by the LIGO detectors during the LIGO-Virgo-KAGRA (LVK) third observing run (O3).
Our search uses a machine learning (ML) based method, Aframe, an alternative to traditional matched filtering search techniques. The O3 observing run has been analyzed by the LVK collaboration, producing GWTC-3, the most recent catalog installment which has been made publicly available in 2021. Various groups outside the LVK have re-analyzed O3 data using both traditional and ML-based approaches. Here, we identify 38 candidates with probability of astrophysical origin ($p_\mathrm{astro}$) greater than 0.5, which were previously reported in GWTC-3. This is comparable to the number of candidates reported by  individual matched-filter searches. In addition, we compare Aframe candidates with catalogs from research groups outside of the LVK, identifying three candidates with $p_\mathrm{astro} > 0.5$. No previously un-reported candidates are identified by Aframe. This work demonstrates that Aframe, and ML based searches more generally, are useful companions to matched filtering pipelines.

\end{abstract}
 \maketitle

\section{Introduction}

The first direct detection of gravitational waves from the merger of a binary stellar-mass black hole system~\cite{AbEA2016a} in 2015 was a breakthrough event in how we observe our universe.
The continuing observations and improved sensitivities of the international network of gravitational-wave detectors (Advanced LIGO~\cite{Aasi_2015}, Advanced Virgo~\cite{Acernese_2015} and KAGRA~\cite{kagra}) over three observing runs (O1, O2 and O3) from 2015-2020, has brought the total number of cataloged systems close to 100~\cite{Abbott_2023}.
Additional candidate events have been reported from non-LVK analyses in the IAS \cite{ias_o3a, ias_o3b}, OGC \cite{3-ogc, 4-ogc}, and AresGW \cite{ares_gw_o3} catalogs, the last of which is another fully machine learning (ML) based approach.
These detections have enabled a broad range of studies, from measuring properties of the binary black hole populations and probing their formation channels to testing general relativity ~\cite{testing-gr, gwtc-3-astro}.
A fourth observing run (O4) of the LVK commenced in May 2023 and is expected to complete in the fall of 2025~\cite{o4_plans}.
LVK data from the first three observing runs are in the public domain through the gravitational-wave open science center~\cite{Abbott_2023_open_data}.


In this work, we perform an end-to-end analysis of the LVK's O3 observing run with Aframe~\cite{aframe_methods}, a machine learning algorithm for detecting binary black hole mergers. We accumulate 100 years of background data by analyzing timeslides across the entire observing run to assign false alarm rates (FARs) to candidates, and perform the same sensitive volume analysis used to assess search pipeline sensitivity in the LVK's GWTC-3. We conduct an injection campaign across O3 to both estimate search sensitivity, and construct a foreground model to estimate candidates probability of astrophysical origin ($p_\mathrm{astro}$). Using $p_\mathrm{astro} > 0.5$ as a threshold for significant candidates, we compare candidates identified by Aframe with the GWTC-3, IAS, OGC, and AresGW catalogs. 


This paper is organized as follows.
In Sec.~\ref{methdat}, we provide a brief review of the Aframe algorithm, the data that was used to train the algorithm, and the data that was analyzed. Sensitivity estimates attained using Monte Carlo techniques are described in section~\ref{sens}.
Our search results are presented in Sec.~\ref{results}, including comparisons with previously published event catalogs.
We conclude with a discussion and outlook for ML searches in Sec.~\ref{sec:conclusion}.
The method used for estimating $p_{\mathrm{astro}}$ is presented in Appendix~\ref{sec:pastro}, and the details of parameter estimation are discussed in Appendix~\ref{sec:pe}.

\section{Methods and Data}
\label{methdat}

Aframe~\cite{Marx_aframe} is a supervised ML framework for training neural networks to detect gravitational-wave signals directly from interferometer strain data. In this work, we use the Aframe configuration described extensively in Ref.~\cite{aframe_methods} to search for gravitational-wave signals from binary black hole (BBH) mergers in O3. 

\begin{figure}[ht]
\centering
\includegraphics[width=\columnwidth]{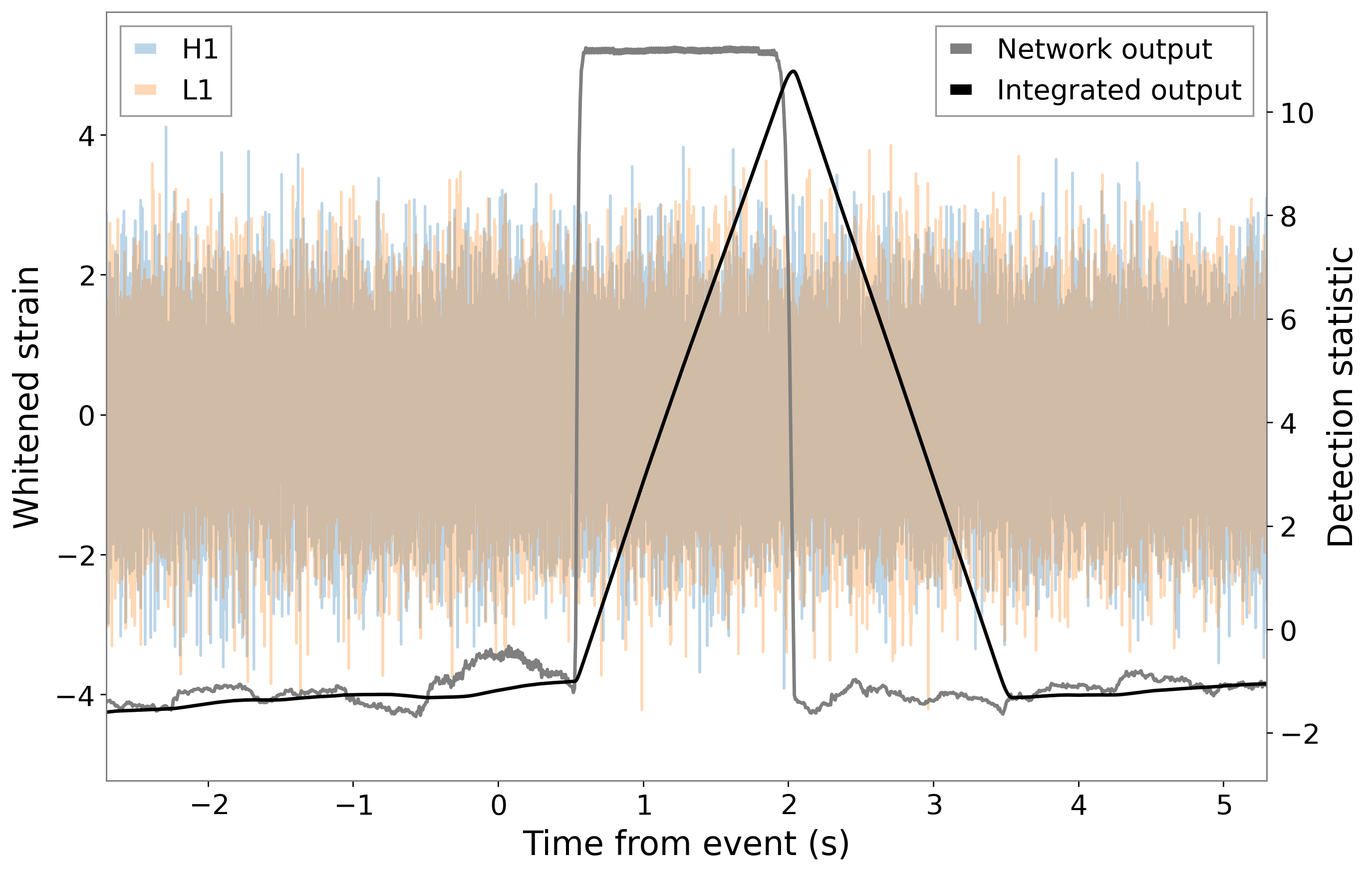}
\caption{
Example of Aframe's neural network output in response to GWTC-3's GW190828\_063405, a merger with $m_1 = 23.7^{+6.8}_{-6.7} M_{\odot}$ and $ m_2 = 10.4^{+3.8}_{-2.2} M_{\odot}$. The event time corresponds to the coalescence time of the binary. As the merger time enters the window of the network, the output increases and remains at a high level while the merger time is within the network window. The integrated response rises as the merger passes through the window, and traditional peak-finding methods can be used to identify events and estimate merger times; see \cite{aframe_methods} for details.
}\label{fig:output_example}
\end{figure}

In brief, the neural network used by Aframe is a ResNet34\cite{resnet}-based architecture that processes overlapping, 1.5 second windows of coincident LIGO Hanford and Livingston strain data to produce a scalar-valued detection statistic. 
Higher values of this statistic signify the neural network is more confident there is a signal consistent with the training distribution present in the data. Aframe analyzes overlapping windows spaced 0.25~s apart, corresponding to an inference sampling rate of 4~Hz. This timeseries of neural-network candidates is integrated using a top hat filter, and candidates are identified by finding peaks in the integrated neural-network output. See Fig.~\ref{fig:output_example} for an illustration of how Aframe's neural network output is used to make detections. 

The neural-network model used in this work has been trained and optimized to search for BBH mergers in the component mass range $[5, 100]~ M_\odot$. The astrophysical priors used to generate the signals used to train the neural-network are reported in Table \ref{table:priors} and were based on the distributions used to estimate search pipeline sensitivity in the LVK's third observing run (O3) \cite{ligo_scientific_collaboration_and_virgo_2023_7890437}. For this analysis, we train and validate Aframe using coincident strain data between times 2019-04-01T:15:00:00 and 2019-04-17T:19:53:20 corresponding to a 17-day calendar period from the start of O3. The training procedure is the same as described in~\cite{aframe_methods}, except here, training is distributed across 8 NVIDIA A100 GPUs on the Nautilus HyperCluster\footnote{https://nrp.ai/}. The learning rate used in~\cite{aframe_methods} is scaled linearly by the number of GPUs to account for the effective increase in batch size from using multiple GPUs~\cite{lr_scaling}.

The search is performed using public data across the entirety of O3 \cite{Buikema_2020, Acernese_2023} provided by the Gravitational Wave Open Science Center (GWOSC) \cite{gwosc}. Currently, Aframe requires coincident Hanford and Livingston data for analysis. For each interferometer, the openly available science mode flag is queried to remove segments with poor data quality. Segments for which the science mode flag is active for both the Hanford and Livingston LIGO interferometers are selected. A minimum coincident segment duration of 128\,s is enforced to allow enough data to accurately estimate the power spectral density. In addition, the LVK provides transient-like veto flags referred to as category 1 and 2 (CAT1, CAT2)~\cite{Abbott_2020, Davis_2021, Virgo:2022fxr, Christensen_2004} vetoes. These binary vetoes are applied to background and foreground events to mitigate the effects of non-Gaussian noise transients. The total live time that meets our analysis criteria is 202.4\,days. The sampling rate of the strain data is downsampled to 2048\,Hz from their original 16384\,Hz, and the strain data is high-passed at 32\,Hz, near the seismic noise wall. 

\section{Sensitivity}
\label{sens}

\begin{figure*}[ht]
\centering
\includegraphics[width=\textwidth]{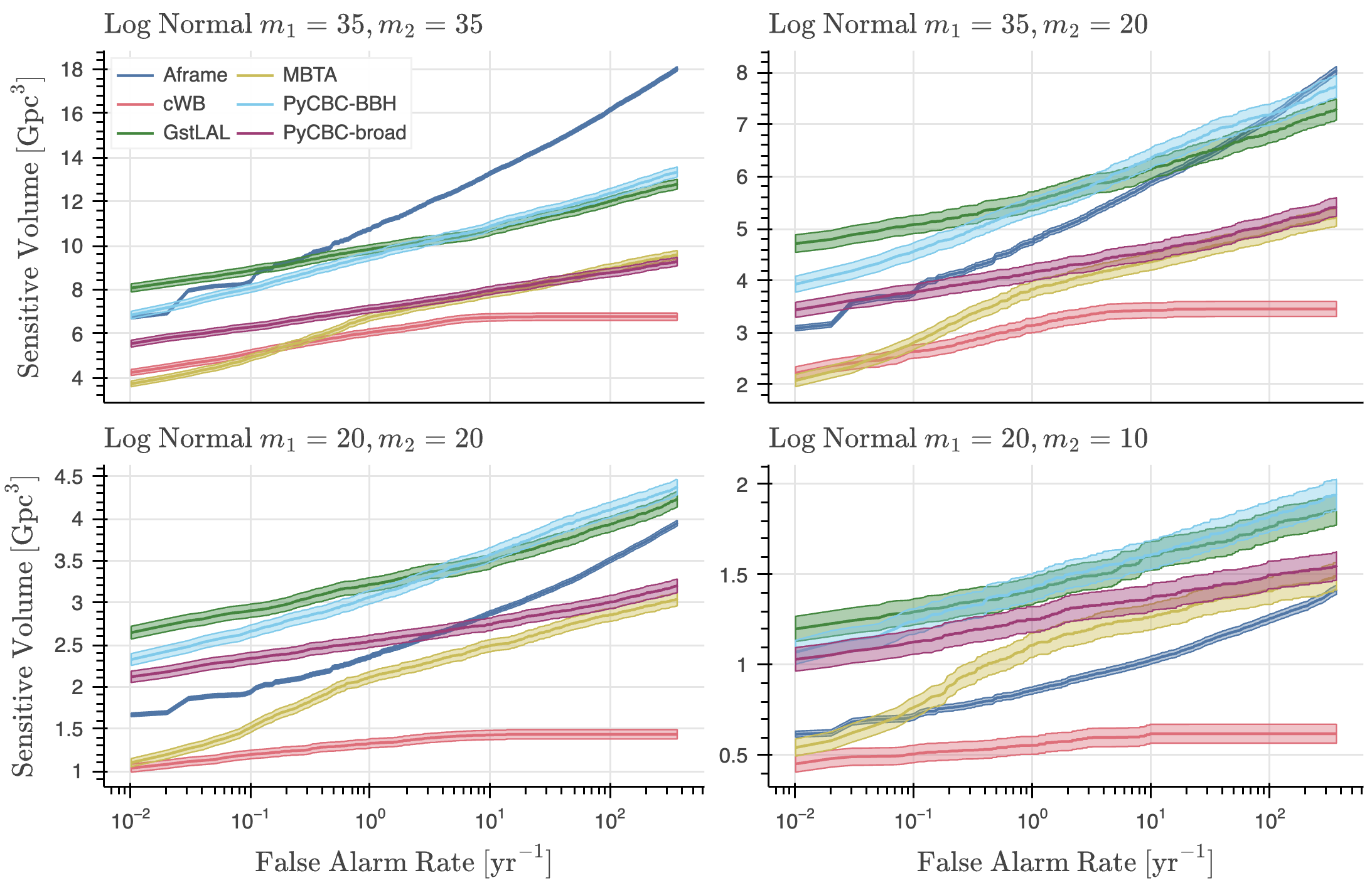}
\caption{Sensitive volume vs FAR for four different mass distributions. Masses are specified in the source frame. Each mass is drawn from a log-normal distribution with a mean of the value given above each plot and a width of 0.1. Aframe demonstrates competitive sensitivity at higher masses, but loses performance relative to traditional search pipelines at lower masses. The sensitive volume of the other pipelines was calculated using data from a GWTC-3 data release\cite{ligo_scientific_collaboration_and_virgo_2023_7890437}.}\label{fig:sensitivity}
\end{figure*}

In Ref.~\cite{aframe_methods}, Aframe's sensitive volume was measured using injections drawn from an astrophysically motivated population, and added into over $\sim$ 1 month of data at the beginning of O3. 
One year of background was accumulated through timeslides and analyzed to assign FARs to injections. 
We refer the reader to Ref.~\cite{aframe_methods} for additional details of that analysis. In this work, sensitive volume is estimated by performing signal injections across O3, excluding the 17-day training period. Using the same data, one-hundred years of timeslides are analyzed to produce a background distribution from which we can assign FARs. Figure~\ref{fig:sensitivity} shows Aframe's sensitive volume for four different mass regions as a function of FAR. Importance sampling is used to estimate sensitive volume for four different mass distributions ~\cite{Tiwari_2018}.
After scaling up the quantity of background analyzed, we see similar trends as reported in ~\cite{aframe_methods}. Aframe achieves similar sensitivity with matched filtering approaches in the 35-35 $M_{\odot}$ bin, and a relative decrease in sensitivity as the chirp mass decreases. Still, this analysis illustrates empirically that Aframe maintains sensitivity at high significances (low FARs).

\input{tables/priors}

\section{Search Results}
The LVK's GWTC-2.1 and GWTC-3 (henceforth, GWTC-3), provided the first catalog of gravitational-wave transients identified during O3. Since then, other analyses have reported additional significant gravitational wave candidates not identified in GWTC-3 \cite{ares_gw_o3, 3-ogc, 4-ogc, ias_o3a, ias_o3b}. As is standard in the gravitational-wave literature when constructing catalogs, we apply an inclusion criteria of $p_{\mathrm{astro}} > 0.5$. $p_{\mathrm{astro}}$ represents the probability that a candidate event has an astrophysical origin (i.e., is not a result of background noise), and is calculated using prior information on the astrophysical population, as well as a search's event distributions for background triggers and simulated event detections. The details of this calculation for Aframe are given in Appendix~\ref{sec:pastro}. Candidates that pass the threshold of 0.5 are called ``significant". Table~\ref{table:significant} lists the 41 significant candidate events that Aframe detects, along with the catalog where the event was first reported. In addition, Table~\ref{table:non_significant} lists 30 candidates identified by Aframe as low-significance, associated with candidates reported in previous catalogs. These low-significant candidates did not pass the $p_{\mathrm{astro}} > 0.5$ threshold, but have a FAR more significant than 2 per year, the threshold for issuing public alerts in O3. Three significant candidates from non-GWTC-3 catalogs are found by Aframe. For these candidates, we perform parameter estimation using Bilby (details in Appendix \ref{sec:pe}). We also refer the reader to recent work \cite{Williams:2024tna} where a configuration similar to that used by the LVK for GWTC-3 was employed to perform parameter estimation on candidates reported from analyses outside the LVK. One event not reported elsewhere was identified with $p_{\mathrm{astro}} = 0.6$, but is discarded due to the identification of a whistle glitch by GravitySpy at the inferred coalescence time. Therefore, Aframe does not identify any previously un-reported candidates. In the rest of this section, we make comparisons of Aframes search results with previously pubished GWTC-3, IAS, OGC, and AresGW catalogs. We note that, when making these comparisons, we include candidates that may lie outside the training prior (Table \ref{table:priors}) used for this search.

\label{results}

\begin{figure}[ht]
\centering
\includegraphics[width=\columnwidth]{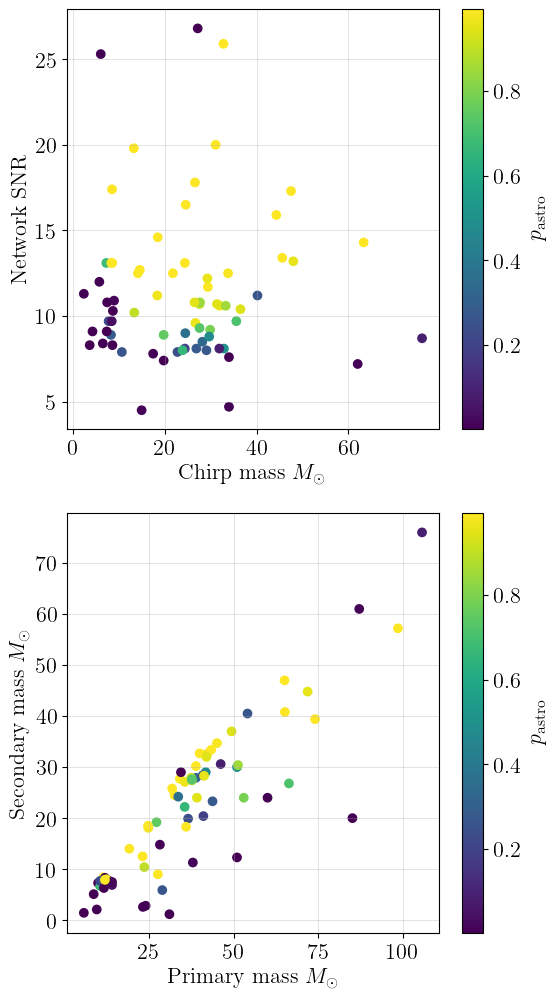}
\caption{\emph{Top}: The SNR vs. the source frame chirp mass of all 70 Hanford/Livingston events reported in GWTC-2.1 and GWTC-3. Mass and SNR information are as reported by the LVK. Points are colored by the $p_{\mathrm{astro}}$ that Aframe reports for each event. Consistent with sensitivity estimates (Fig.~\ref{fig:sensitivity}), Aframe more confidently identifies higher mass candidates. We note that the highest SNR event in this figure, GW200129\_065458, is not detected due to the application of a category 2 veto. \emph{Bottom}: Same events plotted based on source frame component masses.}
\label{fig:missed_found}
\end{figure}

\textbf{\emph{GWTC-3}} 70 candidates were reported in GWTC-3 from O3 with Hanford and Livingston data available. Four matched-filtering pipelines, GstLAL \cite{ewing2023performance, Tsukada_2023}, MBTA \cite{Aubin_2021}, PyCBC-broad, PyCBC-BBH \cite{Dal_Canton_2021}, and one minimally modeled pipeline, cWB \cite{DRAGO2021100678}, reported candidates. 
The criterion for inclusion in GWTC-3 was that any search pipeline reported a $p_{\mathrm{astro}} > 0.5$. Thus, the criterion that only Aframe identifies an event with $p_{\mathrm{astro}} > 0.5$ is more conservative. Table \ref{table:significant} contains 38 of the 70 Hanford/Livingston events reported in GWTC-3 that were also identified by Aframe as significant.
For comparison, GstLAL, MBTA, PyCBC-Broad, PyCBC-BBH, and cWB  detect 47, 48, 41, 61, and 25 events at this threshold respectively \cite{gwtc-2.1, Abbott_2023}. So, Aframe identifies a comparable quantity of candidates. Of the 38 significant Aframe candidates, 27 were detected with FAR more significant than all events in the 100 years of analyzed background, corresponding to $p_{\mathrm{astro}} > 0.99$. In addition, in Table \ref{table:non_significant} we report 16 GWTC-3 candidates which were found by Aframe with low significance. Therefore, there are 16 Hanford/Livingston candidates reported in GWTC-3 that were not identified by Aframe at any significance. One of these candidates, $\mathrm{GW}200129\_065458$ was detected with $p_\mathrm{astro} = 0.99$, but was removed in post processing by a category 2 veto. Of the remaining 15 events, 5 had secondary masses below 5 $M_\odot$, putting them outside of our training prior. The remaining 10 had network matched filter SNRs reported in GWTC-3 below 12.6.
Fig.~\ref{fig:missed_found} summarizes Aframes detections of GWTC-3 events across slices of the parameter space. 

\textbf{\emph{IAS}} The IAS algorithm is another matched-filtering-based detection pipeline. IAS identified a total of 65 significant Hanford/Livingston candidates, 16 of which were reported for the first time by IAS \cite{ias_o3a, ias_o3b}. Of the 65 significant IAS candidates, Aframe identifies 35 as significant, leaving 30 IAS candidates unidentified by Aframe. Six Aframe significant candidates are not identified by IAS. Table \ref{table:significant} lists two candidates, $\mathrm{GW}190707\_083226$ and $\mathrm{GW}190818\_232544$ first identified in the IAS catalog which were also identified as significant by Aframe. In addition, Table \ref{table:non_significant} lists 3 candidates first reported by the IAS analysis which were found by Aframe with low-significance. One such candidate, $\mathrm{GW}200109\_195634$, is found just below threshold with a $p_\mathrm{astro}$ of $0.44$. For this candidate, although the IAS catalog reports a much more significant $p_\mathrm{astro}$, Aframe and the IAS report comparable FARs. We perform parameter estimation for the two significant candidates. Both $\mathrm{GW}190707\_083226$, and $\mathrm{GW}190818\_232544$ have primary mass posterior support in the upper mass gap ($~\sim45-135 M_\odot$) predicted by pulsational pair instability and pair instability supernovae \cite{Woosley_2021}. $\mathrm{GW}190707\_083226$, has effective spin consistent with zero, consistent with the bulk of the GWTC-3 population \cite{gwtc-3-astro}. But, $\mathrm{GW}190818\_232544$ confidently has $\chi_{\mathrm{eff}} \gtrsim 0.5$. 
\newline

\textbf{\emph{OGC}} The 3-OGC and 4-OGC (henceforth OGC) catalogs identify 67 significant Hanford/Livingston candidates, 10 of which were reported for the first time by the OGC \cite{3-ogc, 4-ogc}. Of the 67 significant OGC candidates, Aframe identifies 39 as significant. Two Aframe candidates are not identified by OGC. Table \ref{table:significant} reports 1 candidate, GW200106\_134123, that was first reported by the OGC which Aframe finds identifies as significant. GW200106\_134123 is found by Aframe just above threshold with $p_\mathrm{astro} = 0.55$ and by the OGC analysis with $p_\mathrm{astro} = 0.69$. However, Aframe detects the event at a more significant FAR of $1.16$ per year compared with the FAR of $16.67$ per year reported by the OGC. We perform parameter estimation for this candidate, finding a primary mass of $44^{+8.63}_{-7.27} M_{\odot}$, secondary mass of $27.6^{+6.88}_{-6.39} M_{\odot}$ and $\chi_{\mathrm{eff}}$ consistent with zero.

\textbf{\emph{AresGW}} AresGW is another machine learning-based detection pipeline \cite{aresgw}, that performed a search over O3 \cite{ares_gw_o3}. For calculating $p_\mathrm{astro}$, AresGW employs three variants of their analysis, and assigns the largest $p_\mathrm{astro}$ to each candidate. Therefore, like GWTC-3, their inclusion criteria is less restrictive than ours. AresGW identified a total of 52 significant candidates, 8 of which were reported for the first time by AresGW. Of these, Aframe identifies 33 as significant, leaving 19 AresGW candidates unidentified by Aframe. Of these 19, 8 are also not reported in any of the GWTC-3, IAS, or OGC catalogs. Four of these AresGW-only candidates, however, are found by Aframe as low significance candidates (see Table \ref{table:non_significant}). The most significant of which, GW190607\_083827 is detected by Aframe with $p_\mathrm{astro} = 0.17$, and AresGW with $p_\mathrm{astro} = 0.99$. However, while the $p_\mathrm{astro}$ estimates differ greatly, the FAR's are in closer agreement. This discrepancy in $p_\mathrm{astro}$ despite similar FAR estimates is likely due to Aframe's lack of inclusion of individual source property information in its $p_{\mathrm{astro}}$ foreground model (see Appendix \ref{sec:pastro}).

\section{Conclusion}\label{sec:conclusion}
In this work, an end to end search is performed over the O3 observing run with Aframe, a machine learning based search pipeline. We show that Aframe is capable of detecting previously reported gravitational waves from BBH sources during O3, identifying 38 events reported in GWTC-3 using an inclusion criteria of $p_{\mathrm{astro}} > 0.5$. Additionally, 2 events reported by the IAS analysis and 1 event by the OGC analysis are also found with $p_{\mathrm{astro}} > 0.5$. Many other events reported by GWTC-3, OGC, AresGW and IAS are identified by Aframe as low significance candidates. 

Still, there are a number of areas for improvement. Aframe's failure to identify candidates like GW190814 with SNR $\sim$ 25, but secondary mass $\lesssim 3 M_\odot$ can be mitigated by expanding the training distribution to include lower masses. Missing events like GW191105\_143521 can be explained by Aframe's lower sensitivity to specific regions of the parameter space (see Fig~\ref{fig:sensitivity}). Researching techniques to improve sensitivity in these regions is critical for broadening the searchable parameter space. For example, choosing a training prior more inspired by matched filtering template banks, which aim to minimize the maximum possible waveform mismatch, have been shown to help mitigate this bias neural network approaches have against lower chirp mass signals \cite{Nagarajan:2025hws}. In addition, a fraction of candidates reported in GWTC-3 are currently inaccessible to Aframe due to their reliance on Virgo data. Being able to incorporate arbitrary detector configurations will improve the quantity of data Aframe is able to analyze and increase the number of potential detections. Finally, the current $p_{\mathrm{astro}}$ model only accounts for detection statistic, and does not incorporate candidate source properties with prior astrophysical information to inform the foreground model. For matched filtering pipelines, source properties are extracted from the template that identified the event, and thus are readily available to inform a $p_{\mathrm{astro}}$ model. In its current state, Aframe does not provide source property information on candidates. However, extremely low-latency machine learning based parameter estimation algorithms being developed \cite{Chatterjee:2024pbj, Dax_2021} can be used in conjunction with Aframe to fold source property information into a $p_{\mathrm{astro}}$ model.
\newline

\section{Acknowledgments}
The authors thank Malina Desai and Christina Reissel for fruitful discussions on the manuscript. All authors acknowledge support from the National Science Foundation (NSF) for award PHY-2117997 for the A3D3 Institute. 
W.B and M.W.C acknowledge support from the National Science Foundation (NSF) with grant numbers PHY-2308862 and PHY-2409481. E.M and E.K acknowledge support from the NSF under award PHY-1764464 and PHY-2309200 to the LIGO Laboratory. E.M and E.K also acknowledge NSF awards PHY-1931469 and PHY-1934700. T.B. acknowledges support from the Research Experiences for Undergraduates grant NSF-2348668. E.B., S.H., and K.K. all acknowledge support from the University of Minnesota's Undergraduate Research Opportunities Program.

This research was undertaken with the support of the LIGO computational clusters.
This material is based upon work supported by the NSF's LIGO Laboratory which is a major facility fully funded by the National Science Foundation. 
This research has made use of data obtained from the Gravitational Wave Open Science Center (www.gw-openscience.org), a service of LIGO Laboratory, the LIGO Scientific Collaboration and the Virgo Collaboration.

 The authors are grateful for the computing resources provided by the Nautilus Hypercluster, which is supported in part by National Science Foundation (NSF) awards CNS-1730158, ACI-1540112, ACI-1541349, OAC-1826967, OAC-2112167, CNS-2100237, CNS-2120019, the University of California Office of the President, and the University of California San Diego's California Institute for Telecommunications and Information Technology/Qualcomm Institute.

\input{tables/significant}
\input{tables/non_significant}

\bibliography{references.bib} 
\bibliographystyle{apsrev}

\begin{appendix} 
\section{Probability of Astrophysical Origin}
\label{sec:pastro}

The probability of astrophysical origin $p_\mathrm{astro}$ is often used instead of FAR for ranking candidates because it incorporates the candidates source property estimates in conjunction with prior information on the astrophysical population. As previously noted, in the GWTC-3, IAS, OGC and AresGW catalogs, candidates are identified as significant if they have $p_\mathrm{astro} > 0.5$.

Most searches utilize a coarse-grained approach to $p_\mathrm{astro}$, summing the probability that the candidate belongs to a specific region of the parameter space. Typically, the regions are defined based on the compact objects involved in the merger, i.e. binary black hole (BBH), binary neutron star (BNS), and neutron star black hole (NSBH) regions, corresponding to the $p_\mathrm{astro}$ decomposition 

\begin{equation}
    p_\mathrm{astro} = p_\mathrm{BBH} + p_\mathrm{BNS} + p_\mathrm{NSBH} = 1 - p_\mathrm{terr}
\end{equation}

where $p_\mathrm{terr}$ represents the probability that the event is of terrestrial origin (i.e., that it is not an astrophysical gravitational-wave event).

The typical method for calculating $p_\mathrm{astro}$, known colloquially as the FGMC method \cite{Farr_2015}, models triggers as a Poisson mixture of noise and signal. Under this model, one can derive

\begin{equation}
    p_\mathrm{astro}(x) = \frac{R_s(x)}{R_s(x) + R_n(x)}
\end{equation}
Where $R_s$ and $R_n$ are the expected rates of signals and noise, respectively, at a given detection statistic $x$.
We estimate $R_n$ empirically from noise event detection statistics constructed from 100 years of timeslides performed across the entire O3 observing run. 
First, we model the noise density at a given detection statistic, $f_n(x)$, with a kernel density estimator (KDE) $h(x)$ and an exponential tail,

\begin{equation}
f_n(x)=
    \begin{cases}
        h(x) & \text{if } x \leq x_0 \\
        h(x_0) e^{-\alpha (x - x_0)} & \text{if } x > x_0
    \end{cases}
\end{equation}

where $\alpha$ is fit to the data. This domain split ensures that $p_\mathrm{astro}$ is a smooth, function of $x$ even at values that lack sufficient statistics to be well-modeled by the KDE. Then,

\begin{equation}
    R_n(x) = \frac{f_n(x) N_b}{T}
\end{equation}

Where $T$ is the total length of timeslide data analyzed and $N_b$ is the total number of background events.

$R_s$ is estimated as
\begin{equation}
    R_s(x) = V(x)r
\end{equation}
Where $r$, the astrophysical rate of mergers, comes from prior knowledge typically determined by a population analysis from a previous observing run. $V(x)$ is the sensitive volume of the search at detection statistic $x$, defined as
\begin{equation}
    V(x) = V_0 \frac{N_s(x)}{N_{\text{tot}}}.
\end{equation}

For the astrophysical rate of mergers $r$ we use the mean value inferred by the LVK's population analysis \cite{O3b_pop} from the $\textsc{Power\ Law + Spline}$ model, which is $31 $ $\mathrm{Gpc^{-3}} \mathrm{yr^{-1}}$ for binary systems with component masses in the $[5, 100]~ M_\odot$ range.
This mass range matches the range used for simulated signals in this analysis.
Now, since this mass range consists only of BBH mergers, we implicitly set $p_\mathrm{BNS} = p_\mathrm{NSBH} = 0$ and so our calculated $p_\mathrm{astro}$  is better described as

\begin{equation}
    p_\mathrm{astro} = p_\mathrm{BBH} = 1 - p_\mathrm{terr}
\end{equation}

Sensitive volume is measured through an injection campaign, in which $N_{\text{tot}}$ simulated signals are injected into background data.
The quantity $V_0$ is the astrophysical volume corresponding to the furthest distance at which signals may be simulated, and $N_s(x)$ is the number of signals detected at a detection statistic $x$. The latter quantity is estimated by constructing a foreground model using another KDE.

\begin{figure*}[ht]

\centering
\includegraphics[width=\textwidth]{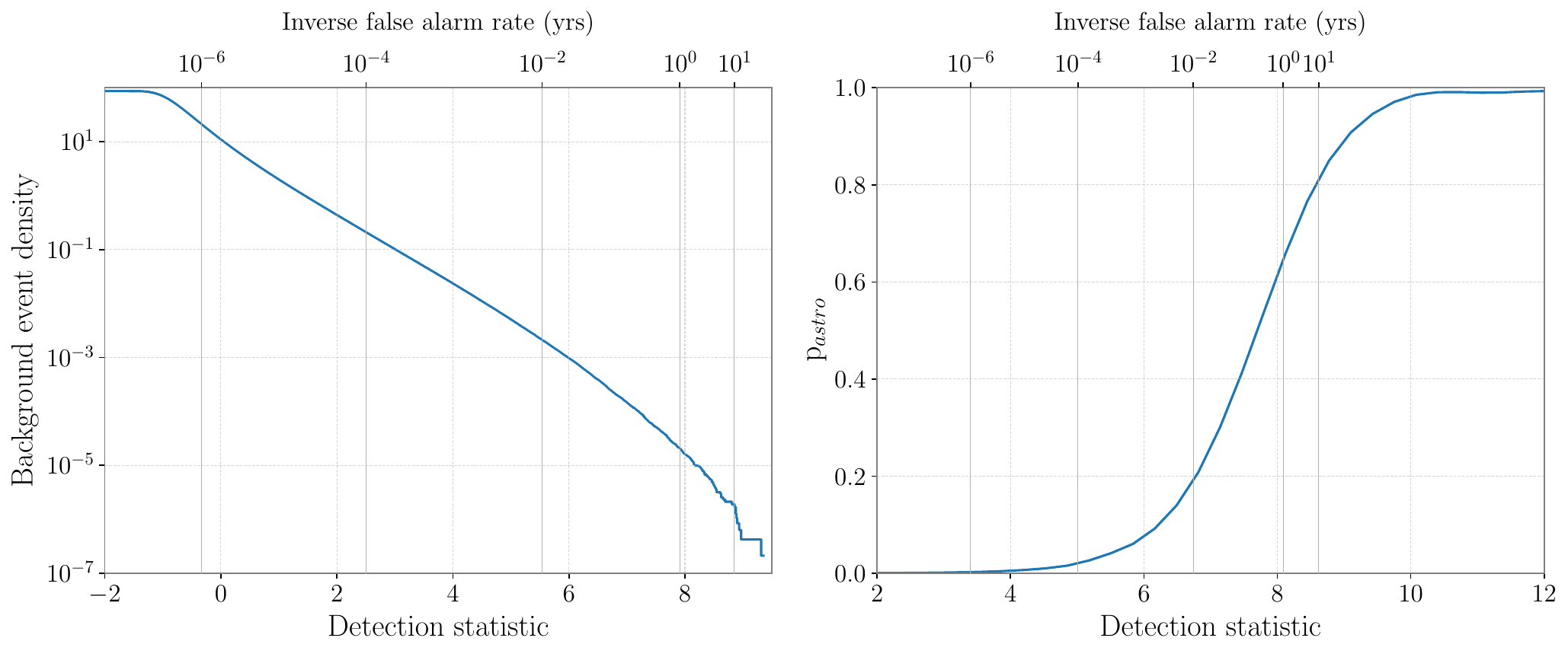}

\caption{\emph{Left}: Cumulative density of detection statistic for background events from 100 years of live time accumulated via the timeslide method. \emph{Right}: $p_\mathrm{astro}$ as a function of Aframe's detection statistic. The top axis of both plots labels the iFAR corresponding to the detection statistic on the bottom axis.}\label{fig:pastro_vs_ifar}
\end{figure*}

Figure \ref{fig:pastro_vs_ifar} shows our $p_\mathrm{astro}$ estimate as a function of detection statistic and inverse FAR (iFAR). Currently, source property information, like component masses, are not incorporated into the foreground model. So, $p_\mathrm{astro}$ is solely a function of our detection statistic, and thus is a smooth, increasing function of detection statistic, or, equivalently, iFAR.

\input{tables/aframe_pe}

\section{Parameter Estimation} \label{sec:pe}
\begin{figure*}
    \centering
    \includegraphics[width=\linewidth]{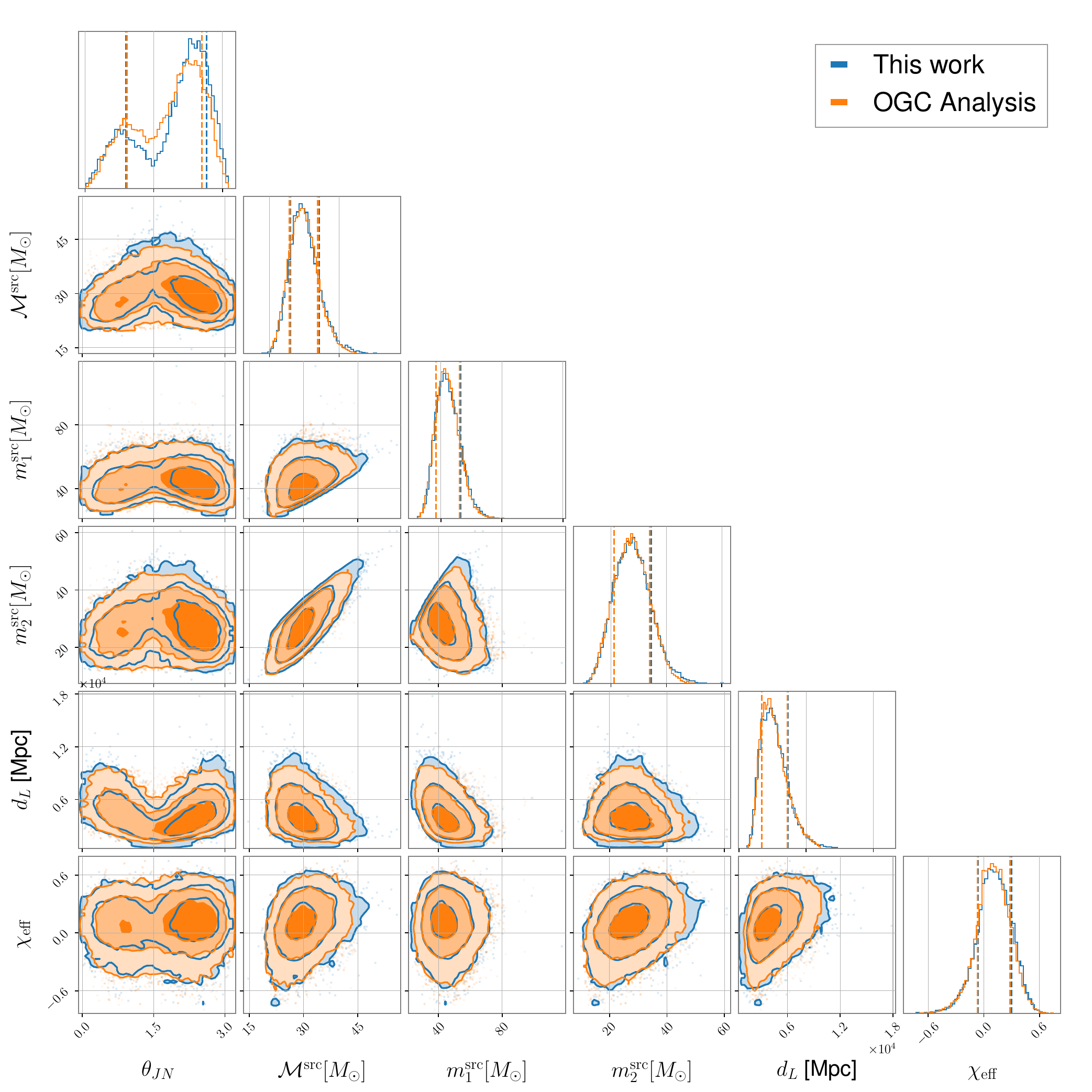}
    \caption{Parameter estimation comparison for OGC event GW200106\_134123}
    \label{fig:gw200106}
\end{figure*}

\begin{figure*}
    \centering
    \includegraphics[width=\linewidth]{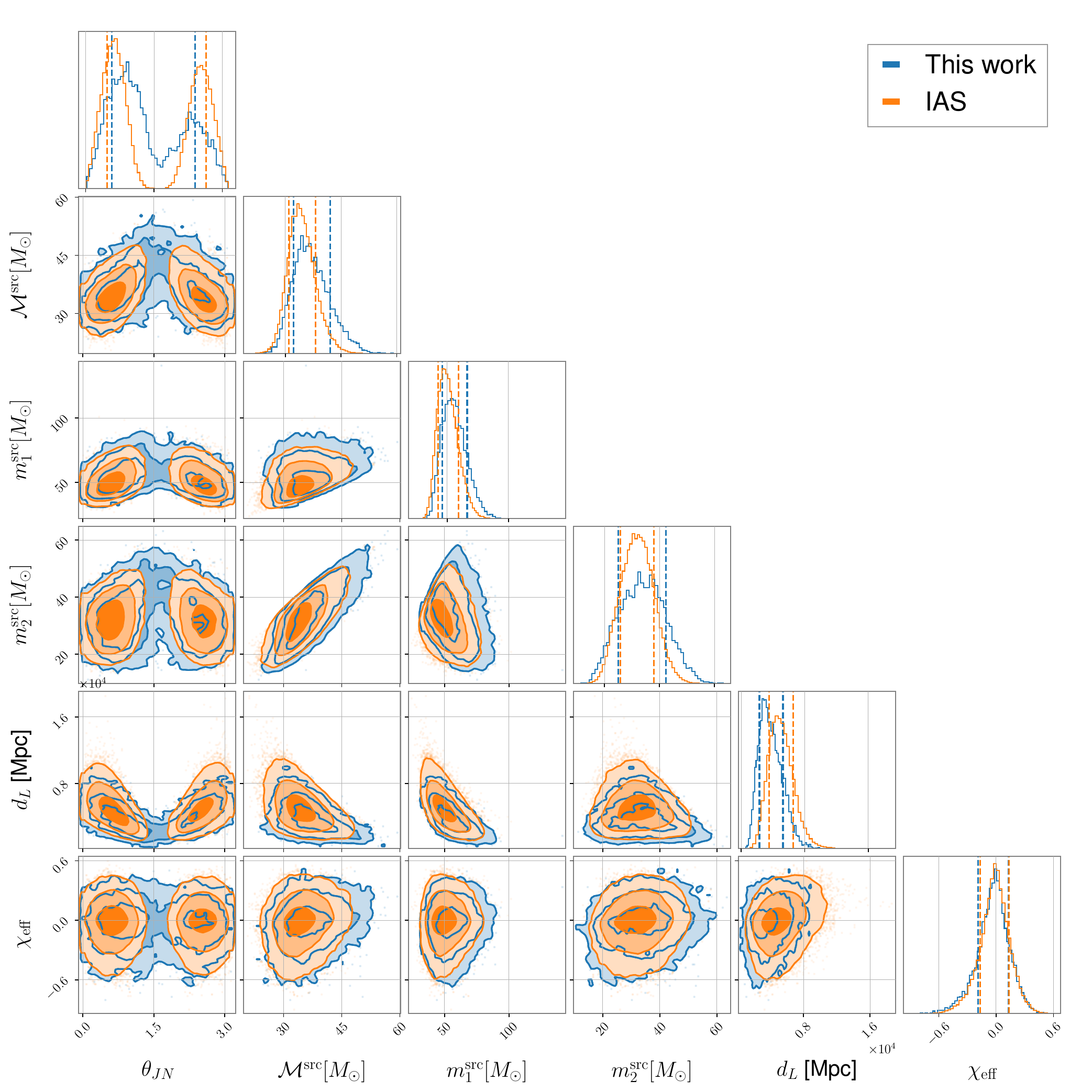}
    \caption{Parameter estimation comparison for IAS event GW190707\_083226}
    \label{fig:gw190707}
\end{figure*}

\begin{figure*}
    \centering
    \includegraphics[width=\linewidth]{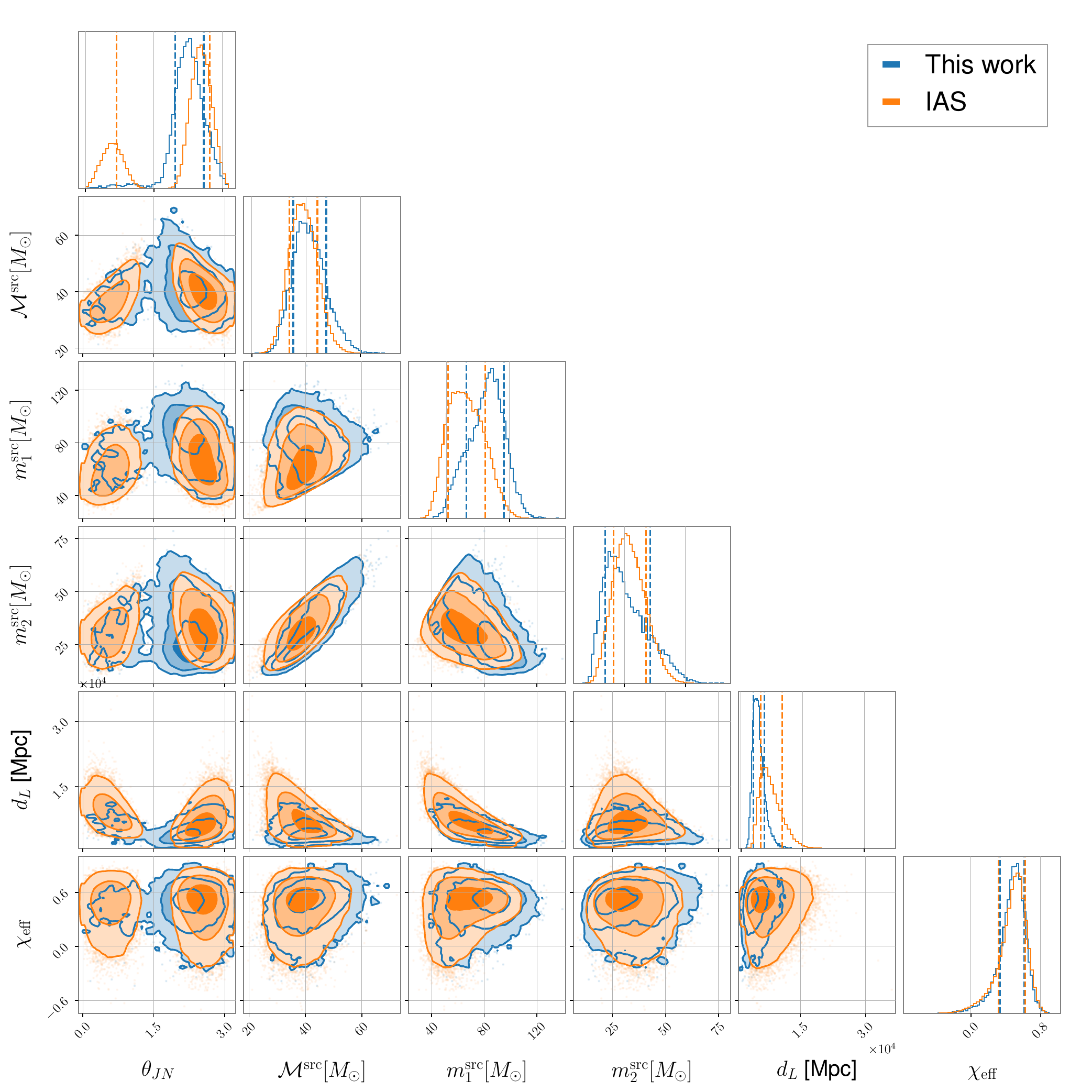}
    \caption{Parameter estimation comparison for IAS event GW190818\_232544}
    \label{fig:gw190818}
\end{figure*}

We perform parameter estimating for the significant candidates Aframe detects that were not analyzed in GWTC-3. These events are GW190707\_083266 and GW190818\_232544 initially reported in the IAS catalog, and GW200106\_134123 initially reported in the OGC catalog. We use Bilby with the IMRPhenomXPHM waveform approximant, and 2000 live points \cite{Ashton_2019}. The Bilby prior distribution used for all events is reported in Fig~\ref{fig:pe-prior}. Estimates for select parameters are listed in Table ~\ref{table:aframe-pe}. Posterior corner plots comparing our anaysis result with those reported by the respective catalog or presented in Figures ~\ref{fig:gw200106},~\ref{fig:gw190707} and ~\ref{fig:gw190818}. The IAS catalog performed two parameter estimation runs with two different spin priors. For a direct comparison, we use their posterior samples from the analysis that used an isotropic spin prior.

For the OGC candidate GW200106\_134123, our parameter estimates agree closely. However, for both IAS candidates GW190818\_232544 and GW190707\_083266, our parameter estimates are broadly consistent, but there are still discrepancies. Most notably, our analyses report smaller luminosity distances compared with the IAS for both events. Despite these differences, for the high $\chi_{\mathrm{eff}}$ candidate GW190818\_232544, our marginalized $\chi_{\mathrm{eff}}$ posteriors are in close agreement.

\begin{figure*}

\begin{verbatim}
mass_1 = bilby.gw.prior.Constraint(name=`mass_1', minimum=5, maximum=200)
mass_2 = bilby.gw.prior.Constraint(name=`mass_2', minimum=5, maximum=200)
mass_ratio = bilby.gw.prior.UniformInComponentsMassRatio(name=`mass_ratio', minimum=0.125, maximum=1)
chirp_mass = bilby.gw.prior.UniformInComponentsChirpMass(name=`chirp_mass', minimum=15, maximum=200)
luminosity_distance = bilby.gw.prior.UniformSourceFrame(name=`luminosity_distance', minimum=1e2, maximum=3.5e4)
dec = Cosine(name=`dec')
ra = Uniform(name=`ra', minimum=0, maximum=2 * np.pi, boundary=`periodic')
theta_jn = Sine(name=`theta_jn')
psi = Uniform(name=`psi', minimum=0, maximum=np.pi, boundary=`periodic')
phase = Uniform(name=`phase', minimum=0, maximum=2 * np.pi, boundary=`periodic')
a_1 = Uniform(name=`a_1', minimum=0, maximum=0.99)
a_2 = Uniform(name=`a_2', minimum=0, maximum=0.99)
tilt_1 = Sine(name=`tilt_1')
tilt_2 = Sine(name=`tilt_2')
phi_12 = Uniform(name=`phi_12', minimum=0, maximum=2 * np.pi, boundary=`periodic')
phi_jl = Uniform(name=`phi_jl', minimum=0, maximum=2 * np.pi, boundary=`periodic')
\end{verbatim}
\caption{Bilby prior distribution used for parameter estimation analyses.} \label{fig:pe-prior}
\end{figure*}

\end{appendix}

\end{document}

%% file: tables/priors.tex
\begin{table}
\centering
\footnotesize
\begin{tabular*}{\columnwidth}{ p{4cm} p{1.5cm} p{1.75cm} p{1.5cm} }
\hline
\hline
Parameter & Prior & Limits & Units\\
\hline
Mass of primary ($m_1$) & $m_1^{-2.35}$ & $[5, 100]$ & $M_{\odot}$\\
Mass of secondary ($m_2$) & $m_2$ & $[5, m_1]$ & $M_{\odot}$\\
Redshift ($z$) & Comoving & $[0, 2]$ & -\\
Polarization angle ($\psi$) & Uniform & $[0, \pi]$ & rad.\\
Spin magnitude ($a_1$, $a_2$) & Uniform & $[0, 0.998]$ & -\\
Spin tilt ($\theta_1$, $\theta_2$) & Sine & $(0, \pi)$ & rad.\\
Spin azimuthal angle ($\phi_{12}$) & Uniform & $[0, 2\pi]$ & rad.\\
Spin phase angle ($\phi_{JL}$) & Uniform & $[0, 2\pi]$ & rad.\\
Orbital phase ($\phi_{c}$) & Uniform & $[0, 2\pi]$ & rad.\\
Right ascension ($\alpha$) & Uniform & $[0, 2\pi]$ & rad.\\
Declination ($\delta$) & Cosine & $[-\pi/2, \pi/2]$ & rad.\\
Inclination angle ($\theta_{JN}$) & Sine & $[0, \pi]$ & rad.\\
\hline
\hline
\end{tabular*}
\caption{Priors on parameters used to generate waveforms for both the training and testing sets. The prior is derived from that used in GWTC-3 to assess search pipeline sensitivity. The component mass distributions are defined in the source frame. ``Comoving" refers to uniform in comoving volume.}
\label{table:priors}
\end{table}

%% file: tables/significant.tex
\begin{table*}[ht]
\centering
\caption{Candidates that Aframe detects with $p_\mathrm{astro} \geq 0.5$. The catalog each candidate was originally reported in is listed. Events marked with an asterisk occurred during the training period. Candidates identified by Aframe with FAR $\leq 0.01$ $\mathrm{yr}^{-1}$ correspond to events that are more significant than the entire 100 years of analyzed background. For candidates first identified in GWTC-3, $m_1$ and $m_2$ are as reported there. For the OGC and IAS events, we report our own $m_1$ and $m_2$ estimates.}
\label{table:significant}
\rowcolors{2}{white}{gray!15}
\begin{tabular}{!{\vrule width 1pt}lcccccccc!{\vrule width 1pt}}
\toprule[1pt]
 & Event & Catalog & Aframe FAR $\mathrm{yr}^{-1}$ & FAR $\mathrm{yr}^{-1}$ & Aframe $p_\mathrm{astro}$ & $p_\mathrm{astro}$ & $m_1^\mathrm{src} (M_{\odot})$ & $m_2^\mathrm{src} (M_{\odot})$ \\
\midrule[1pt]
1 & GW190412\_053044$^{*}$ & GWTC-3 & $\leq 0.01$ & $1.00 \times 10^{-5}$ & 0.99 & 1.00 & $27.7^{+6.0}_{-6.0}$ & $9.0^{+2.0}_{-1.4}$ \\
2 & GW191204\_171526 & GWTC-3 & $\leq 0.01$ & $1.00 \times 10^{-5}$ & 0.99 & 0.99 & $11.7^{+3.3}_{-1.7}$ & $8.4^{+1.3}_{-1.7}$ \\
3 & GW190728\_064510 & GWTC-3 & $\leq 0.01$ & $1.00 \times 10^{-5}$ & 0.99 & 1.00 & $12.5^{+6.9}_{-2.3}$ & $8.0^{+1.7}_{-2.6}$ \\
4 & GW200311\_115853 & GWTC-3 & $\leq 0.01$ & $1.00 \times 10^{-5}$ & 0.99 & 0.99 & $34.2^{+6.4}_{-3.8}$ & $27.7^{+4.1}_{-5.9}$ \\
5 & GW191109\_010717 & GWTC-3 & $\leq 0.01$ & $1.80 \times 10^{-4}$ & 0.99 & 0.99 & $65.0^{+11.0}_{-11.0}$ & $47.0^{+15.0}_{-13.0}$ \\
6 & GW200224\_222234 & GWTC-3 & $\leq 0.01$ & $1.00 \times 10^{-5}$ & 0.99 & 0.99 & $40.0^{+6.7}_{-4.5}$ & $32.7^{+4.8}_{-7.2}$ \\
7 & GW190519\_153544 & GWTC-3 & $\leq 0.01$ & $1.00 \times 10^{-5}$ & 0.99 & 1.00 & $65.1^{+10.8}_{-11.0}$ & $40.8^{+11.5}_{-12.7}$ \\
8 & GW190521\_074359 & GWTC-3 & $\leq 0.01$ & $1.00 \times 10^{-5}$ & 0.99 & 1.00 & $43.4^{+5.8}_{-5.5}$ & $33.4^{+5.2}_{-6.8}$ \\
9 & GW190706\_222641 & GWTC-3 & $\leq 0.01$ & $5.00 \times 10^{-5}$ & 0.99 & 1.00 & $74.0^{+20.1}_{-16.9}$ & $39.4^{+18.4}_{-15.4}$ \\
10 & GW190521\_030229 & GWTC-3 & $\leq 0.01$ & $1.30 \times 10^{-3}$ & 0.99 & 1.00 & $98.4^{+33.6}_{-21.7}$ & $57.2^{+27.1}_{-30.1}$ \\
11 & GW190513\_205428 & GWTC-3 & $\leq 0.01$ & $1.30 \times 10^{-5}$ & 0.99 & 1.00 & $36.0^{+10.6}_{-9.7}$ & $18.3^{+7.4}_{-4.7}$ \\
12 & GW190915\_235702 & GWTC-3 & $\leq 0.01$ & $1.00 \times 10^{-5}$ & 0.99 & 1.00 & $32.6^{+8.8}_{-4.9}$ & $24.5^{+4.9}_{-5.8}$ \\
13 & GW200225\_060421 & GWTC-3 & $\leq 0.01$ & $1.10 \times 10^{-5}$ & 0.99 & 0.99 & $19.3^{+5.0}_{-3.0}$ & $14.0^{+2.8}_{-3.5}$ \\
14 & GW190408\_181802$^{*}$ & GWTC-3 & $\leq 0.01$ & $1.00 \times 10^{-5}$ & 0.99 & 1.00 & $24.8^{+5.4}_{-3.5}$ & $18.5^{+3.3}_{-4.0}$ \\
15 & GW190828\_063405 & GWTC-3 & $\leq 0.01$ & $1.00 \times 10^{-5}$ & 0.99 & 1.00 & $31.9^{+5.4}_{-4.1}$ & $25.8^{+4.9}_{-5.3}$ \\
16 & GW190707\_093326 & GWTC-3 & $\leq 0.01$ & $1.00 \times 10^{-5}$ & 0.99 & 1.00 & $12.1^{+2.6}_{-2.0}$ & $7.9^{+1.6}_{-1.3}$ \\
17 & GW190727\_060333 & GWTC-3 & $\leq 0.01$ & $1.00 \times 10^{-5}$ & 0.99 & 1.00 & $38.9^{+8.9}_{-6.0}$ & $30.2^{+6.5}_{-8.3}$ \\
18 & GW190512\_180714 & GWTC-3 & $\leq 0.01$ & $1.00 \times 10^{-5}$ & 0.99 & 1.00 & $23.2^{+5.6}_{-5.6}$ & $12.5^{+3.5}_{-2.6}$ \\
19 & GW191222\_033537 & GWTC-3 & $\leq 0.01$ & $1.00 \times 10^{-5}$ & 0.98 & 0.99 & $45.1^{+10.9}_{-8.0}$ & $34.7^{+9.3}_{-10.5}$ \\
20 & GW191215\_223052 & GWTC-3 & $\leq 0.01$ & $1.00 \times 10^{-5}$ & 0.97 & 0.99 & $24.9^{+7.1}_{-4.1}$ & $18.1^{+3.8}_{-4.1}$ \\
21 & GW200128\_022011 & GWTC-3 & $\leq 0.01$ & $4.30 \times 10^{-3}$ & 0.97 & 0.99 & $42.2^{+11.6}_{-8.1}$ & $32.6^{+9.5}_{-9.2}$ \\
22 & GW200209\_085452 & GWTC-3 & $\leq 0.01$ & 0.05 & 0.96 & 0.97 & $35.6^{+10.5}_{-6.8}$ & $27.1^{+7.8}_{-7.8}$ \\
23 & GW190503\_185404 & GWTC-3 & $\leq 0.01$ & $1.00 \times 10^{-5}$ & 0.96 & 1.00 & $41.3^{+10.3}_{-7.7}$ & $28.3^{+7.5}_{-9.2}$ \\
24 & GW190602\_175927 & GWTC-3 & $\leq 0.01$ & $1.00 \times 10^{-5}$ & 0.96 & 1.00 & $71.8^{+18.1}_{-14.6}$ & $44.8^{+15.5}_{-19.6}$ \\
25 & GW190517\_055101 & GWTC-3 & $\leq 0.01$ & $3.50 \times 10^{-4}$ & 0.95 & 1.00 & $39.2^{+13.9}_{-9.2}$ & $24.0^{+7.4}_{-7.9}$ \\
26 & GW190421\_213856 & GWTC-3 & $\leq 0.01$ & $2.80 \times 10^{-3}$ & 0.95 & 1.00 & $42.0^{+10.1}_{-7.4}$ & $32.0^{+8.3}_{-9.8}$ \\
27 & GW191230\_180458 & GWTC-3 & $\leq 0.01$ & 0.05 & 0.94 & 0.96 & $49.4^{+14.0}_{-9.6}$ & $37.0^{+11.0}_{-12.0}$ \\
28 & GW200219\_094415 & GWTC-3 & 0.02 & $9.90 \times 10^{-4}$ & 0.91 & 0.99 & $37.5^{+10.1}_{-6.9}$ & $27.9^{+7.4}_{-8.4}$ \\
29 & GW190828\_065509 & GWTC-3 & 0.02 & $3.50 \times 10^{-5}$ & 0.90 & 1.00 & $23.7^{+6.8}_{-6.7}$ & $10.4^{+3.8}_{-2.2}$ \\
30 & GW190413\_134308$^{*}$ & GWTC-3 & 0.10 & 0.18 & 0.85 & 0.99 & $51.3^{+16.6}_{-12.6}$ & $30.4^{+11.7}_{-12.7}$ \\
31 & GW200208\_130117 & GWTC-3 & 0.10 & $3.10 \times 10^{-4}$ & 0.85 & 0.99 & $37.7^{+9.3}_{-6.2}$ & $27.4^{+6.3}_{-7.3}$ \\
32 & GW191127\_050227 & GWTC-3 & 0.18 & 0.25 & 0.79 & 0.74 & $53.0^{+47.0}_{-20.0}$ & $24.0^{+17.0}_{-14.0}$ \\
33 & GW191204\_110529 & GWTC-3 & 0.28 & 3.30 & 0.75 & 0.74 & $27.3^{+10.8}_{-5.9}$ & $19.2^{+5.5}_{-6.0}$ \\
34 & GW190803\_022701 & GWTC-3 & 0.35 & 0.07 & 0.73 & 0.97 & $37.7^{+9.8}_{-6.7}$ & $27.6^{+7.6}_{-8.5}$ \\
35 & GW190929\_012149 & GWTC-3 & 0.42 & 0.16 & 0.71 & 0.87 & $66.3^{+21.6}_{-16.6}$ & $26.8^{+14.7}_{-10.6}$ \\
36 & GW191129\_134029 & GWTC-3 & 0.48 & $1.00 \times 10^{-5}$ & 0.68 & 0.99 & $10.7^{+4.1}_{-2.1}$ & $6.7^{+1.5}_{-1.7}$ \\
37 & GW190527\_092055 & GWTC-3 & 0.56 & 0.23 & 0.66 & 0.85 & $35.6^{+18.7}_{-8.0}$ & $22.2^{+9.0}_{-8.7}$ \\
38 & GW200106\_134123 & OGC & 1.16 & 16.67 & 0.55 & 0.69 & ${44.2}_{-7.3}^{+8.6}$ & ${27.6}_{-6.4}^{+6.9}$ \\
39 & GW190707\_083226 & IAS & 1.34 & 0.04 & 0.52 & 0.94 & ${55.2}_{-9.3}^{+11.1}$ & ${33.9}_{-8.8}^{+8.4}$ \\
40 & GW190818\_232544 & IAS & 1.47 & 0.29 & 0.51 & 0.81 & ${82.7}_{-16.9}^{+12.8}$ & ${28.8}_{-8.3}^{+13.8}$ \\
41 & GW200216\_220804 & GWTC-3 & 1.52 & 0.35 & 0.51 & 0.77 & $51.0^{+22.0}_{-13.0}$ & $30.0^{+14.0}_{-16.0}$ \\
\bottomrule[1pt]
\end{tabular}

\end{table*}

%% file: tables/non_significant.tex
\begin{table*}[ht]
\centering
\caption{Candidates that Aframe detects with $p_\mathrm{astro} < 0.5$. The catalog each candidate was originally reported in is listed. Events marked with an asterisk occurred during the training period. Parameter estimates for $m_1^\mathrm{src}$ and $m_2^\mathrm{src}$, are as reported in the respective catalog.}
\label{table:non_significant}
\rowcolors{2}{white}{gray!15}
\begin{tabular}{!{\vrule width 1pt}lcccccccc!{\vrule width 1pt}}
\toprule[1pt]
 & Event & Catalog & Aframe FAR $\mathrm{yr}^{-1}$ & FAR $\mathrm{yr}^{-1}$ & Aframe $p_\mathrm{astro}$ & $p_\mathrm{astro}$ & $m_1^\mathrm{src} (M_{\odot})$ & $m_2^\mathrm{src}(M_{\odot})$ \\
\midrule[1pt]
1 & GW190731\_140936 & GWTC-3 & 1.89 & 0.33 & 0.47 & 0.83 & $41.8^{+12.7}_{-9.1}$ & $29.0^{+10.2}_{-9.9}$ \\
2 & GW200109\_195634 & IAS & 2.29 & 1.08 & 0.44 & 0.81 & $69_{-19}^{+24}$ & $48_{-17}^{+22}$ \\
3 & GW190413\_052954$^{*}$ & GWTC-3 & 3.45 & 0.82 & 0.36 & 0.93 & $33.7^{+10.4}_{-6.4}$ & $24.2^{+6.5}_{-7.0}$ \\
4 & GW200220\_124850 & GWTC-3 & 4.22 & 30.00 & 0.34 & 0.83 & $38.9^{+14.1}_{-8.6}$ & $27.9^{+9.2}_{-9.0}$ \\
5 & GW191103\_012549 & GWTC-3 & 5.74 & 0.46 & 0.29 & 0.94 & $11.8^{+6.2}_{-2.2}$ & $7.9^{+1.7}_{-2.4}$ \\
6 & GW190916\_200658 & OGC & 5.91 & 4.55 & 0.28 & 0.88 & $45.7^{+17.0}_{-12.3}$ & $24.0^{+13.2}_{-10.4}$ \\
7 & GW190514\_065416 & GWTC-3 & 6.23 & 2.80 & 0.27 & 0.76 & $40.9^{+17.3}_{-9.3}$ & $28.4^{+10.0}_{-10.1}$ \\
8 & GW191113\_071753 & GWTC-3 & 6.59 & 26.00 & 0.26 & 0.68 & $29.0^{+12.0}_{-14.0}$ & $5.9^{+4.4}_{-1.3}$ \\
9 & GW190701\_203306 & GWTC-3 & 6.63 & $5.70 \times 10^{-3}$ & 0.26 & 1.00 & $54.1^{+12.6}_{-8.0}$ & $40.5^{+8.7}_{-12.1}$ \\
10 & GW190711\_030756 & IAS & 7.65 & 0.09 & 0.24 & 0.93 & $80_{-40}^{+50}$ & $18_{-7}^{+11}$ \\
11 & GW190719\_215514 & GWTC-3 & 7.94 & 0.63 & 0.24 & 0.92 & $36.6^{+42.1}_{-11.1}$ & $19.9^{+10.0}_{-9.3}$ \\
12 & GW191105\_143521 & GWTC-3 & 9.32 & 0.01 & 0.21 & 0.99 & $10.7^{+3.7}_{-1.6}$ & $7.7^{+1.4}_{-1.9}$ \\
13 & GW190926\_050336 & OGC & 11.10 & 3.70 & 0.19 & 0.88 & $40.1^{+19.1}_{-10.4}$ & $23.4^{++10.8}_{-9.2}$ \\
14 & GW190607\_083827 & AresGW & 14.39 & 6.50 & 0.17 & 0.99 & $40.5_{+12.0}^{-7.6}$ & $31.0_{+9.1}^{-8.2}$ \\
15 & GW200214\_223307 & OGC & 15.88 & 12.50 & 0.16 & 0.72 & $51.6^{+24.4}_{-14.2}$ & $30.9^{+15.5}_{-12.0}$ \\
16 & GW200129\_114245 & OGC & 20.52 & 25.00 & 0.13 & 0.53 & $79.1^{+40.2}_{-37.6}$ & $31.5^{+18.6}_{-14.4}$ \\
17 & GW190805\_211137 & GWTC-3 & 39.57 & 0.63 & 0.08 & 0.95 & $46.2^{+15.4}_{-11.2}$ & $30.6^{+11.8}_{-11.3}$ \\
18 & GW190426\_190642 & GWTC-3 & 41.06 & 4.10 & 0.08 & 0.75 & $105.5^{+45.3}_{-24.1}$ & $76.0^{+26.2}_{-36.5}$ \\
19 & GW190426\_082124 & AresGW & 50.19 & 20.00 & 0.07 & 0.50 & $31.5_{+22.5}^{-11.3}$ & $13.8_{+6.9}^{-5.2}$ \\
20 & GW200305\_084739 & OGC & 85.79 & 50.00 & 0.05 & 0.59 & $33.8^{+12.1}_{-7.7}$ & $23.2^{+7.7}_{-9.4}$ \\
21 & GW200318\_191337 & OGC & 89.64 & 2.00 & 0.05 & 0.97 & $49.1^{+16.4}_{-12.0}$ & $31.6^{+12.0}_{-11.6}$ \\
22 & GW200225\_075134 & IAS & 141.55 & 6.67 & 0.03 & 0.60 & $51_{-11}^{+17}$ & $37_{-11}^{+13}$ \\
23 & GW190930\_133541 & GWTC-3 & 145.58 & 0.01 & 0.03 & 1.00 & $14.2^{+8.0}_{-4.0}$ & $6.9^{+2.4}_{-2.1}$ \\
24 & GW190705\_164632 & AresGW & 156.73 & 49.00 & 0.03 & 0.51 & $44.7_{+24.8}^{-12.8}$ & $23.0_{+11.7}^{-9.8}$ \\
25 & GW190906\_054335 & IAS & 156.99 & 1.37 & 0.03 & 0.61 & $37_{-8}^{+12}$ & $24_{-8}^{+8}$ \\
26 & GW190720\_000836 & GWTC-3 & 361.50 & $1.00 \times 10^{-5}$ & 0.01 & 1.00 & $14.2^{+5.6}_{-3.3}$ & $7.5^{+2.2}_{-1.8}$ \\
27 & GW190614\_134749 & AresGW & 454.01 & 4.60 & 0.01 & 0.99 & $37.0_{+31.8}^{-10.7}$ & $25.2_{+15.2}^{-9.7}$ \\
28 & GW200322\_091133 & GWTC-3 & 586.91 & 140.00 & 0.01 & 0.62 & $38.0^{+130.0}_{-22.0}$ & $11.3^{+24.3}_{-6.0}$ \\
\bottomrule[1pt]
\end{tabular}

\end{table*}

%% file: tables/aframe_pe.tex
\begin{table*}[ht]
\centering
\caption{Bilby parameter estimates for the three significant candidates identified by Aframe that were not reported in GWTC-3}
\label{table:aframe-pe}
\rowcolors{2}{white}{gray!15}
\begin{tabular}{!{\vrule width 1pt}lcccccc!{\vrule width 1pt}}
\toprule[1pt]
& Event & $\mathcal{M}_c^\mathrm{src} (M_\odot)$ & ${m}_1^\mathrm{src} (M_\odot)$ & ${m}_2^{src} (M_\odot)$ & $D_L$ (Gpc) & $\chi_\mathrm{eff}$ \\
\midrule[1pt]
1 & GW200106\_134123 & ${29.8}_{-3.7}^{+4.5}$ & ${44.2}_{-7.3}^{+8.6}$ & ${27.6}_{-6.4}^{+6.9}$ & ${4.06}_{-1.35}^{+1.79}$ & ${0.12}_{-0.18}^{+0.19}$ \\
2 & GW190707\_083226 & ${36.5}_{-4.3}^{+5.5}$ & ${55.2}_{-9.3}^{+11.1}$ & ${33.9}_{-8.8}^{+8.4}$ & ${3.53}_{-1.25}^{+1.72}$ & ${-0.02}_{-0.17}^{+0.15}$ \\
3 & GW190818\_232544 & ${41.0}_{-5.7}^{+6.5}$ & ${82.7}_{-16.9}^{+12.8}$ & ${28.8}_{-8.3}^{+13.8}$ & ${4.09}_{-1.12}^{+1.56}$ & ${0.49}_{-0.16}^{+0.12}$ \\
\bottomrule[1pt]
\end{tabular}
\end{table*}